\documentclass[12pt]{iopart}

\usepackage{graphicx}
\usepackage{dcolumn}
\usepackage{bm}
\usepackage{amssymb}
\usepackage{epsf}
\usepackage{color}
\usepackage[utf8]{inputenc}

\begin{document}

\title[]{Non-equilibrium magnetic response in concentrated spin-glass AuFe(11\%) alloy}

\author{Sudip Pal and S. B. Roy}
\address{
 UGC DAE Consortium for Scientific Research\\
 University Campus, Khandwa Road\\
 Indore-452001, India 
}

\author{A. K. Nigam\footnote[2]{presently retired from TIFR, Mumbai}}
\address{
 Tata Institute of Fundamental Research\\
 Mumbai-400005, India 
}

\ead{sudip.pal111@gmail.com,sindhunilbroy@gmail.com}
%----------------------------------------------------------------------------------------------------------------------------------------------------------------------------------------------------------------------------
\begin{abstract}
We report a detailed study of dc magnetization and ac susceptibility performed on the zero field cooled ($ZFC$) and field cooled ($FC$) state of polycrystalline AuFe(11\%) alloy. The temperature variation of $ZFC$ and $FC$ dc magnetization at low fields show a distinct peak around $T_f$ = 33 K, which indicates the cooperative freezing of the finite size spin clusters.  A weak thermomagnetic irreversibility between $ZFC$ and $FC$ magnetization appears at a temperature $T_{ir}$, which is slightly below $T_f$.  Further down the temperature the $ZFC$ magnetization curve shows a weak shoulder at $T_{sh} < T_f$, where the $ZFC$ and $FC$ magnetization curves show strong bifurcation indicating the onset of strong thermomagnetic irreversibility. With the increase in the external magnetic field, $T_{ir}$ drops faster than the freezing temperature $T_f$, which becomes indistinguishable at higher fields. The $ZFC$ ac susceptibility shows a sharp cusp at $T_f$, which shifts towards higher temperatures with an increase in the frequency of the ac magnetic field. When ac susceptibility is recorded after cooling the sample from high temperature in the presence of dc bias magnetic field, the susceptibility cusp gets broadened. In this case, no perceptible frequency dependency of the $T_f$ has been observed, but a significant dispersion in the ac susceptibility is present below $T_f$. This clearly indicates the non-equilibrium nature of the $FC$ state. The $ZFC$ state of the AuFe(11\%) alloy exhibits memory effect, which indicates the metastable nature of the $ZFC$ state. In addition, the FC state of AuFe(11\%) alloy also exhibits a pronounced memory effect which further underlines that the FC state is not an equilibrium state. In contrast to the general perception obtained through the mean-field theories of thermodynamic phase transition in spin-glass envisaging the $FC$ state to be an equilibrium state, the present experimental results clearly indicate that the energy landscape of the $FC$ state of AuFe(11\%) alloy is a nontrivial one. 
\end{abstract}      
                       
%----------------------------------------------------------------------------------------------------------------------------------------------------------------------------------------------------------------------------
\section{Introduction} 
Magnetic materials in presence of competing interactions and quenched disorder often exhibit time and history-dependent physical properties. Intense research activities are going on over decades to gain an understanding of such systems, which in turn has been applied to understand various complex natural events {\color{blue}\cite{Sadati2014,Keim2019,Morgan2020}}. In this direction, the binary alloys, such as AuMn, AgMn, AuFe, etc. popularly known as $canonical$ $spin-glasses$ have served as model systems over many decades. In these alloys, transition metal atoms with a local magnetic moment in the dilute limit are randomly distributed in noble metal hosts such as Ag, Au, Cu etc. These local moments interact with each other through long-range Ruderman–Kittel–Kasuya–Yosida (RKKY) interaction and gives rise to a spin-glass state, where individual magnetic moments freeze in random directions. Starting from the dilute limit, with the increase in the concentration of transition metal atoms, instead of individual spins, small cluster of spins with short-range magnetic order forms. These spin clusters then act as a giant spin and freeze in a random direction below a characteristic freezing temperature. Such concentrated alloys are variously termed in literature as cluster spin-glass, concentrated spin-glass, mictomagnet, etc. Above a certain concentration, known as percolation threshold, an infinite cluster of transition metal atoms forms and a long-range magnetic order is established at least in a part of the system {\color{blue}\cite{Coles1978,Sarkissian1981,Binder1986,Mydosh1993,Mydosh2015}}. 

The zero field cooled ($ZFC$) state of the canonical spin-glasses has been investigated with various macroscopic as well as microscopic experimental techniques like dc magnetization, ac susceptibility, specific heat, neutron scattering, M$\ddot{o}$sbauer spectroscopy, muon spin resonance studies, etc. {\color{blue}\cite{Binder1986,Mydosh1993,Mydosh2015}}. This $ZFC$ state is found to be a non-equilibrium state with time and history dependent properties {\color{blue}\cite{Binder1986,Mydosh1993,Mydosh2015}}. These interesting physical properties are understood within the mean-field theoretical framework involving  Parisi’s solution of the infinite-range Sherrington-Kirkpatrick (SK) model that predicts a thermodynamic second-order phase transition from the high-temperature paramagnetic state to the low-temperature spin-glass state {\color{blue}\cite{sherrington,parisi}}. Within the same theoretical framework, cooling the system
in the presence of a finite magnetic field from a paramagnetic state would bring the spin-glass to an “infinite time equilibrium state” {\color{blue}\cite{Binder1986,Mydosh1993,Mydosh2015}}. This is the field cooled ($FC$) state of spin-glass, and it is widely believed to be an equilibrium state of the system. However, in recent times we have shown that the $FC$ state in dilute canonical spin-glass systems like AuMn(1.8\%) and AgMn(1.1\%) alloys is not really an equilibrium state of the system {\color{blue}\cite{Sudip1, Sudip2}}. The $FC$ state of these canonical spin-glasses clearly shows the well-known magnetic memory effect in dc magnetization studies and frequency dependence in ac-susceptibility measurements {\color{blue}\cite{Sudip1, Sudip2}}. 

In Ising spin glasses, the spin-glass transition temperature in the $H-T$ plane obeys de Almeida Thouless (AT) line {\color{blue}\cite{dealmeida}}. In Heisenberg spin-glass systems with weak random anisotropy, in the presence of the external magnetic field, the transverse and the longitudinal components of the spins may behave differently {\color{blue}\cite{courtenay,kenning}} . When the system is cooled at a constant magnetic field, the transverse spin component undergoes a spin freezing temperature which is called as Gabay-Toulouse (G-T) critical line {\color{blue}\cite{gabay}}. On the other hand, at a further lower temperature, there is a cross-over line where the longitudinal spin component freezes. This lower temperature cross-over line follows similar power law dependence as AT line in the $H-T$ plane {\color{blue}\cite{kenning,fisher}}.  It is indeed interesting to note here that two kinds of irreversible regimes - weak and strong - have been observed experimentally in concentrated or cluster spin-glass systems. For example upon lowering the temperature at the constant magnetic field ( $H >$ 500 Oe) in a concentrated spin-glass system Cu$_{0.94}$Mn$_{0.06}$, one first encounter a weak irreversibility regime at a characteristic temperature, which represents a true phase transition with transverse freezing. This weak irreversibility follows the G-T relation {\color{blue}\cite{gabay}} with a coefficient close to the mean-field value for a Heisenberg system {\color{blue}\cite{kenning}}. Then at a further lower temperature, there is an onset of strong irreversibility, which is found to agree quantitatively with the mean-field d'Almeida-Thouless-like H-T crossover relation for longitudinal freezing in a Heisenberg system {\color{blue}\cite{kenning}}. In AuFe(5\%) spin-glass alloy, using torque and magnetization measurements, the transverse component of the spins has been found to freeze at a higher temperature than the longitudinal spin component {\color{blue}\cite{petit}}

The question now naturally arises what would be the behaviour of the $FC$ state in the concentrated or cluster spin-glass systems? To this end, to the best of our knowledge, there is no detailed information available in the literature on the $FC$ state of canonical cluster spin-glasses. In this backdrop, we have made an investigation of the bulk magnetic properties of a polycrystalline sample of AuFe(11\%) alloy both in the $ZFC$ and $FC$ state of the sample. At this Fe concentration, AuFe system is well below the percolation limit of long-range ferromagnetic order around 15\% Fe, and lies in the region, which is marked as cluster spin-glass in the Au-Fe magnetic phase diagram {\color{blue}\cite{Sarkissian1981}}. We have performed detailed dc magnetization ($M$) and ac susceptibility ($\chi$) measurements in the temperature ($T$) range of 2-300 K, and up to 70 kOe  magnetic of field ($H$). Our results among other things reveal a distinct non-equilibrium behaviour in the $FC$ state of this cluster spin-glass system.  

%-----------------------------------------------------------------------------------------------------------
\begin{figure*}[t]
\centering
\includegraphics[scale=0.48]{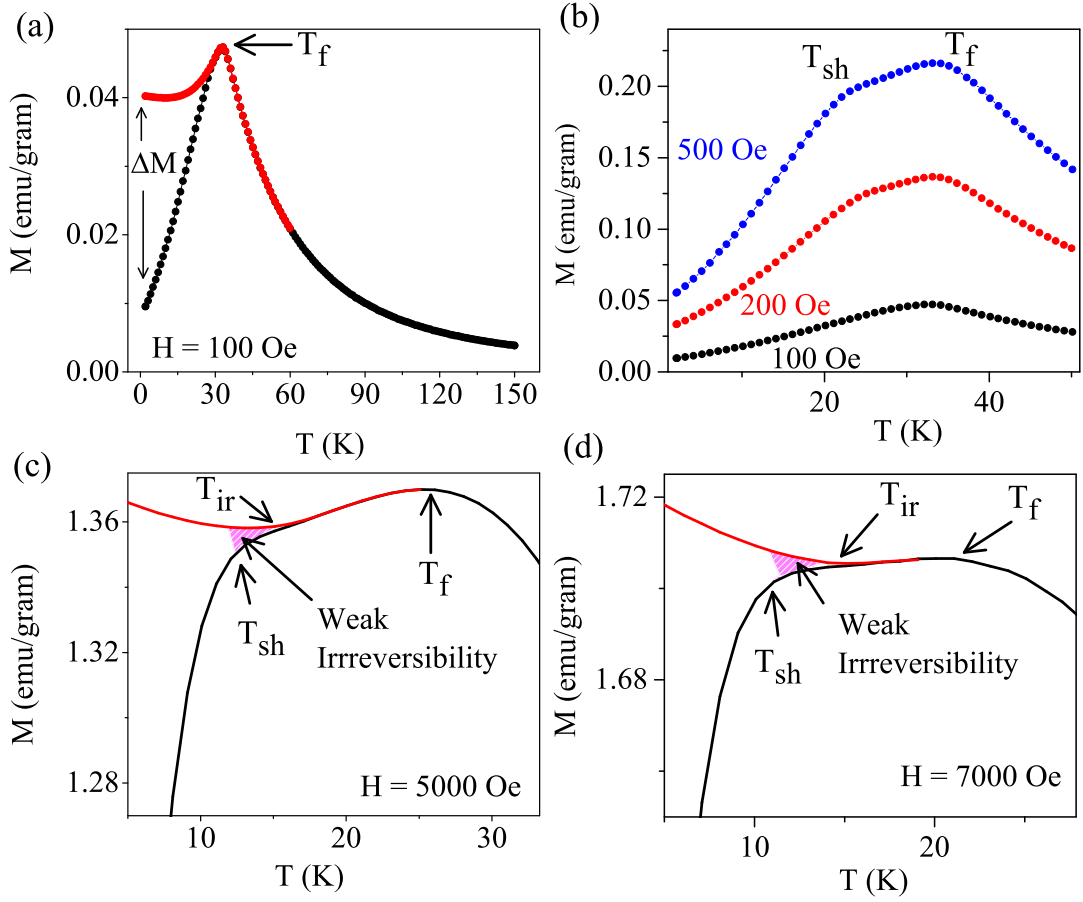}
\caption{(a) Magnetization ($M$) versus temperature ($T$) plot  of AuFe(11\%) alloy  at 100 Oe obtained under $ZFC$ and $FCW$ protocols; (b) $ZFC$ magnetization curves in a magnified scale at H = 100, 300 and 500 Oe to reveal a shoulder in magnetization at a temperature $T_{sh} < T_f$.  (c) and (d) presents the $M-T$ plot  of AuFe(11\%) alloy at $H$ = 5 and 7 kOe, respectively, highlighting weak irreversibility regime $T_{ir} > T > T_{sh}$ and strong irreversibility regime $T < T_{sh}$ . }
\end{figure*}

\begin{figure}[t]
\centering
\includegraphics[scale=0.46]{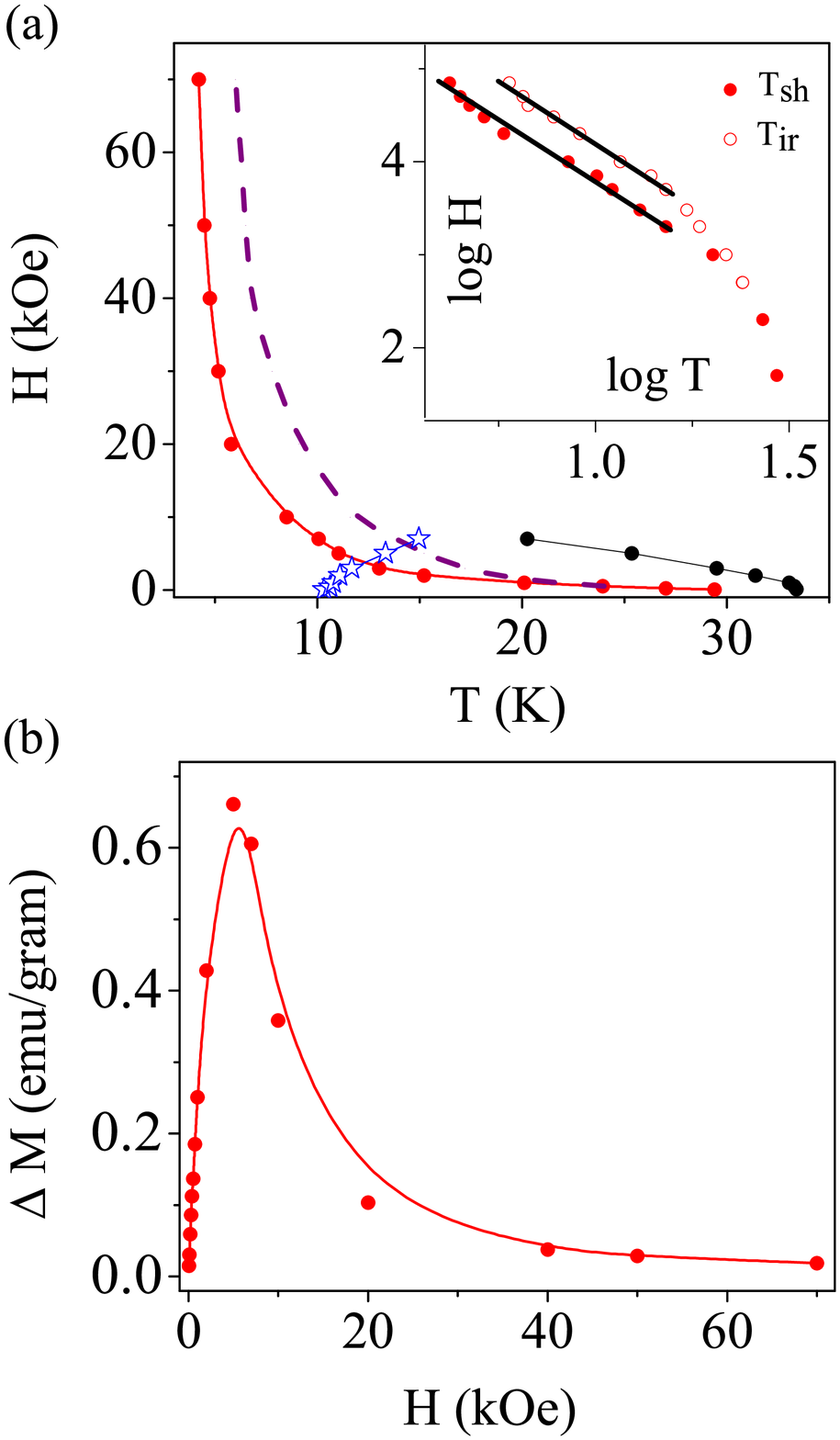}
\caption{(a)  Variation of the characteristic temperatures  $T_{f}$ $T_{ir}$ and $T_{sh}$,  $T_{min}$ of AuFe(11\%) alloy with magnetic field,  obtained from the $ZFC$ and $FCW$ magnetization curves at different fields. Black filled circle, red filled circles and blue star represent $T_f$, $T_{sh}$ and $T_{min}$, respectively. The dotted line indicates the characteristic temperature $T_{ir}$ where weak thermomagnetic irreversibility between $ZFC$ and $FCW$ magnetization is observed. The inset of (a) presents double logarithmic plot of the $T_{sh}$,  $T_{ir}$ and $H$. (b) Variation of thermomagnetic irreversibility $\Delta M$ of AuFe(11\%) alloy at T = 2 K with magnetic field up to 70 kOe. The solid line is the guide line to the eye.}
\end{figure}

%----------------------------------------------------------------------------------------------------------------------------------------------------------------------------------------------------------------------------
\section{Experimental details} 
The AuFe(11\%) sample has been prepared by arc melting method  {\color{blue}\cite{nigam}}. Magnetic measurements have been carried out in MPMS3, SQUID-VSM (M/S Quantum design, USA).  Temperature dependent magnetization $M$ shown in Fig. {\color{blue}1(a)} and {\color{blue}1(b)} has been measured in the $ZFC$ and field cooled warming ($FCW$) protocols. In $ZFC$ protocol, the sample is cooled down to the lowest temperature (here 2 K) in the absence of any external magnetic field ($H$). At $T$ = 2 K, the measurement field is applied and $M-T$ curve is recorded during warming. After recording $M$ in the $ZFC$ mode, the sample is subsequently cooled down to $T$ = 2 K at the same field and temperature dependence of $M$ has been measured during warming to obtain the $FCW$ magnetization. The temperature sweep rate during the measurements of $M$ is 1 K/min. The $ZFC$ ac susceptibility has been measured after the sample has been cooled down to $T$ = 2 K in the absence of an external dc magnetic field and data have been recorded during warming at an ac magnetic field of $H_{ac}$ = 1 Oe. In case of $FC$ ac-susceptibility measurements, the sample has been cooled to $T$ = 2 K in the presence of a certain dc magnetic field $H_{dc}$ and the data have been recorded during warming with an ac magnetic field $H_{ac}$ = 1 Oe while keeping the dc bias field $H_{dc}$ on. The ac-susceptibility at each temperature has been recorded after stabilizing the temperature for each data point. 

%----------------------------------------------------------------------------------------------------------------------------------------------------------------------------------------------------------------------------
\section{Results and discussion}
 
\subsection{Spin-glass freezing, thermomagnetic irreversibility and characteristic temperatures} 
In Fig. {\color{blue}1(a)}, we have shown the temperature dependence of $M$ in the $ZFC$ and $FCW$ modes at $H$ = 100 Oe of AuFe(11\%) alloy.  Magnetization shows a distinct peak at $T_f$ = 32 K in both $ZFC$ and $FCW$ modes, which corresponds to the freezing temperature associated with the transition from high temperature paramagnetic to cluster spin-glass state. The value of $T_f$ is consistent with the phase diagram obtained earlier using the magnetic susceptibility and neutron scattering measurements {\color{blue}\cite{Sarkissian1981,Cannella1972}}.  Below $T_f$ the $ZFC$ magnetization starts to decrease, and then at a characteristic temperature $T_{sh}$ magnetization  drops relatively fast after exhibiting a weak shoulder. In the $FCW$ mode starting from 2 K magnetization first decreases with the increase in temperature and then reaches a shallow minimum around 10 K before starting to rise and ultimately reaching a peak at $T_f$. In this FC protocol, $M(T)$ does not reveal any significant structure around $T_{sh}$. 

Above the freezing temperature $T_f$, the $ZFC$ and $FCW$ magnetization curves are completely merged and resemble a paramagnetic-like behaviour. However, it should be mentioned that susceptibility deviates from the Curie-Weiss like behavior expected from a purely paramagnetic state at much higher temperatures than $T_f$. The bifurcation between the ZFC and FC magnetization namely $themomagnetic$ $irreversibility$ appears at a temperature $T_{ir}$, which is slightly below the peak temperature of magnetization  $T_f$.  We define the thermomagnetic irreversibility as $\Delta$M = ($M_{FCW} - M_{ZFC}$), and we note that it remains relatively small below $T_{ir}$ until the temperature reaches $T_{sh}$.  where  $\Delta M$ increases abruptly. We recall here that at $T_{sh}$, $ZFC$ magnetization undergoes a change in the slope and exhibits a shoulder. We will elaborate below more on these two distinct irreversibility regions, i.e., a weak irreversibility below $T_{ir}$ and strong irreversibility below $T_{sh}$. However, at low magnetic fields, the characteristic temperatures $T_{ir}$ and $T_{sh}$ nearly coincide,  and in our present sample these two irreversibility regions can be clearly distinguished approximately above $H$ = 2 kOe.  

The temperature dependence of $M$ in both the $ZFC$ and $FCW$ modes change significantly with the increase in the applied magnetic field. The temperature $T_f$, where $M$ in the $ZFC$ and $FCW$ curves show a peak gradually shifts towards lower temperatures and the peak itself broadens with the increase in the field. More significantly, the shoulder in $M$ observed in the $ZFC$ mode at $T_{sh}$ becomes progressively more pronounced. This can be clearly observed in Fig. {\color{blue}1(b)}, where we have shown the $ZFC$ magnetization curves at $H$ = 100, 300 and 500 Oe. Note that at $H$ = 100 Oe, M exhibits a clear peak at $T_f$, and a weak shoulder at a lower temperature is barely visible. As the measurement field is increased, the shoulder at $T_{sh}$ becomes more pronounced and it moves towards lower temperatures. An interesting effect of the field is that the $T_{sh}$ reduces at a faster rate compared to $T_f$ and hence the temperature span ($T_f - T_{sh}$) gradually expands. 

The observation of the two different characteristic temperatures $T_{ir}$ and $T_{sh}$ associated with weak and strong irreversibilities, respectively, is interesting. In Fig. {\color{blue}1(c)} and {\color{blue}1(d)}, we have shown the expanded view of the $M-T$ response at $H$ = 5 and 7 kOe to highlight these two irreversibility regions more clearly. Below $T_{sh}$, the $ZFC$ magnetization sharply drops and a strong irreversibility is observed. But, a weak irreversibility continues to exist until $T_{ir}$ which lies distinctly above T$_{sh}$. In Fig. {\color{blue}2(a)}, we have shown the variation of $T_f$, $T_{sh}$ and $T_{ir}$ with applied magnetic field.  The observation of weak and strong irreversibilities below $T_{ir}$ and $T_{sh}$, respetively, in AuFe(11\%) can prima facie be associated with the freezing of the transverse and longitudinal spin components respectively. According to the mean field theory of spin-glass, the G-T and the A-T lines vary in $H-T$ plane following $H^{\alpha} \propto (1-\frac{T_g(H)}{T_g(0)})$, with $\alpha$ = 2 and 2$/$3 respectively {\color{blue}\cite{dealmeida,courtenay,kenning,gabay,fisher,petit}}. Here, $T_g(0)$ and $T_g(H)$ are the freezing temperatures at zero and finite fields $\cite{gabay}$. Therefore, the log-log plot of H and $T_{ir}$ and $T_{sh}$ should be linear in our  AuFe(11\%) sample, where the slope of the linear variation should give the value of $\alpha$. In the inset of Fig. {\color{blue}2(a)}, we have shown the double logarithmic plot of the $T_{sh}$, $T_{ir}$ and H for AuFe(11\%) sample. It may be noted that these two temperatures are merged at low fields, and the log-log plot is linear only at higher fields. The linear fitting of the data in this range gives the value of the slope approximately $\alpha$ = 0.38 for both $T_{sh}-H$, and $T_{ir}-H$ lines, which is different from the expectation of mean field theory of spin-glass {\color{blue}\cite{dealmeida,courtenay,kenning,gabay,fisher,petit}}. In addition, in Fig. {\color{blue}2(a)} we have presented the field dependence of the $T_{min}$, where the $FCW$ magnetization curve shows a minima. $T_{min}$ increases with increase in the field. However, at higher magnetic fields, a clear minimum in the $FCW$ magnetization curve is no longer visible.

The dependence of the thermomagnetic irreversibility, $\Delta M$  with the applied magnetic field is also non-trivial. In Fig. {\color{blue}2(b)}, we have shown the variation of $\Delta M$ at $T$ = 2 K, which exhibits various interesting features:
\begin{enumerate}
\item $\Delta$ M  varies non-monotonically; it initially increases and exhibits maximum around $H_{max}$ = 5 kOe, and reduces again at higher fields.
\item $\Delta M$ varies non-linearly with $H$ at both sides of $H_{max}$, and overall it takes an asymmetric bell like shape.
\item $\Delta M$ increases faster below $H_{max}$ and decays very slowly at higher fields. Finite and significant $\Delta M$ exits even at our highest measurement field of $H$ = 70 kOe.
\end{enumerate}

%----------------------------------------------------------------------------------------------------------

\begin{figure*}[t]
\centering
\includegraphics[scale=0.46]{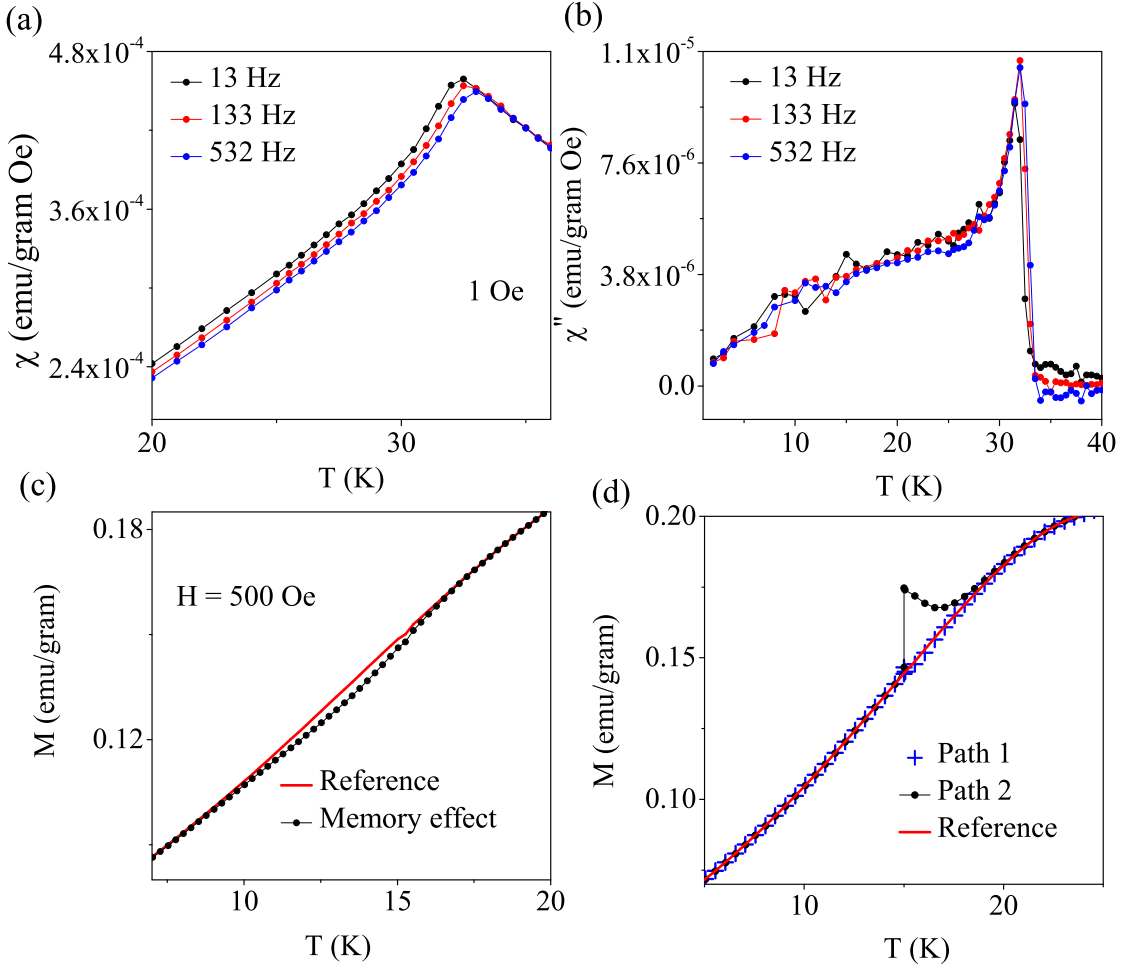}
\caption{(a) Real and  (b) imaginary part of the ac susceptibility (c) memory effect in the $ZFC$ state (d) effect of the magnetic field on the memory effect in the $ZFC$ State of AuFe(11\%) alloy . The reference curve is the usual $ZFC$ magnetization curve at $H$ = 500 Oe. Path 1 corresponds to the measurement protocol where $H$ is reduced from $H$ = 500 to 300 Oe. Path 2 corresponds to the measurement protocol where $H$ is raised from $H$ = 500 to 700 Oe.  }
\end{figure*}

%-----------------------------------------------------------------------------------------------------------
\subsection{Metastable zero field cooled state} 
In Fig. {\color{blue}3(a)} and {\color{blue}3(b)}, we have shown the temperature dependence of the real ($\chi^{\prime}$) and imaginary part ($\chi^{\prime\prime}$) of the ac susceptibility of  AuFe(11\%) alloy at different frequencies which has been recorded in the ZFC state with an ac magnetic field $H_{ac}$ = 1 Oe. The $\chi^{\prime}$ of AuFe(11\%) (see Fig. {\color{blue}3(a)}) shows a prominent cusp at $T_f$ = 33.5 K at the frequency of $f$ = 13 Hz. Above $T_f$ the $\chi^{\prime}$ - $T$ plots at different frequencies are entirely merged which is expected in the paramagnetic state of the system. However, finite dispersion in the $\chi^{\prime}$ at different frequencies exists at $T_f$, which persist down to the lowest measured temperature of T = 2 K. It may also be noted that with the increase in frequency the peak position $T_f$ progressively shifts towards higher temperature. In principle, the frequency-dependent ac susceptibility probes the system in the time scale of $1/f$ and is a sensitive tool to characterize the magnetic state of these alloys systems {\color{blue}\cite{Sarkissian1981}}. The frequency dependence of the ac susceptibility in the probing frequency range implies the inherent slow dynamics and distribution of the time scale of the magnetic entities in the system which are frozen below $T_f$. It has been reported earlier that above the percolation concentration $C_P$, as the temperature is reduced from high temperature paramagnetic state, $\chi^{\prime}$ shows a sharp peak at $T_C$ which is an indication of the transition from paramagnet to long-range ferromagnetic state {\color{blue}\cite{Sarkissian1981}}. At lower concentration of the magnetic ions, near to $C_P$, $\chi^{\prime}$ shows a relatively broad peak corresponding to quasicritical behavior which is associated to ferromagnetic ordering within a large but not infinite cluster. In addition, smaller clusters present in the system may also undergo a freezing transition at a further lower temperature. Therefore, the temperature dependence of $\chi^{\prime}$ shows a double peak at concentrations slightly lower than $C_P$. In our case, the AuFe (11\%) lies distinctly below this percolation limit and represents a concentrated spin-glass phase. Here, in addition to this sharp cusp at $T_f$, an additional broad hump-like feature is observed at low temperatures. This feature is very feeble in the $\chi^{\prime}$, but more clearly visible rather in the imaginary part of the ac susceptibility ($\chi^{\prime\prime}$), shown in Fig. {\color{blue}3(b)}. Above $T_f$, $\chi^{\prime\prime}$ is absent, which is expected in the paramagnetic state. As the temperature is reduced, $\chi^{\prime\prime}$ shows a sharp rise at $T_f$ and then decreases after exhibiting a cusp. This is followed by a broad and prominent hump-like feature at lower temperatures.

The frozen $ZFC$ spin-glass state exhibits clear signatures of metastability also in the dc magnetization measurements. One example is the observation of memory effect, where the system can store a memory of the previous events below $T_f$ and that reflects in the magnetization measurements. This can be realized following different measurement protocols, one of which has been shown in Fig. {\color{blue}3(c)}. In this protocol, the sample is first cooled down to $T$ = 2 K in absence of external magnetic field and a reference magnetization curve is recorded during warming at H = 500 Oe. It may be noted that this reference curve is identical to the $M-T$ response of the system measured in $ZFC$ mode at $H$ = 500 Oe. To check the memory effect, the sample is then cooled at $H$ = 0 down to $T_m$ =  15 K, where cooling is paused for a time period of $t_w$ = 1 hour. Then, the temperature is again reduced down to T = 2 K and M is recorded during warming after applying the field of H = 500 Oe. Note that, these two magnetization curves are completely merged in the entire temperature region except around $T_m$ = 15 K, where the memory curve shows a shallow dip and the reference and the memory curve has small dispersion. This indicates that the system can memorize the temperature pause at $T_m$ performed in the cooling path. This is known as the memory effect and is a prominent signature of the metastable nature of the $ZFC$ state of this cluster spin-glass  AuFe(11\%) .

%----------------------------------------------------------------------------------------------
\begin{figure*}[t]
\centering
\includegraphics[scale=0.31]{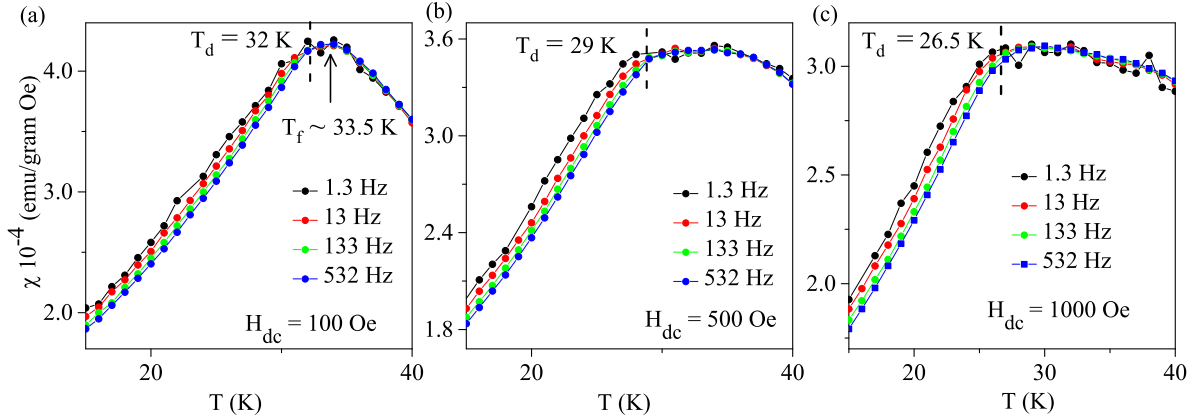}
\caption{The real part of the ac susceptibility of AuFe(11\%) alloy the $FC$ state in presence of dc magnetic field: (a) $H_{dc}$ = 100, (b) 500 and (c) 1000 Oe. The $FC$ state was prepared after cooling the sample in the same dc bias magnetic field from the high temperature paramagnetic state to 2 K.}
\end{figure*}
%----------------------------------------------------------------------------------------------

In Fig. {\color{blue}3(d)}, we have shown the memory effect following a different measurement protocol, which probes the effect of the magnetic field on the $ZFC$ state of the system. In this case, we have first measured the reference magnetization curve at $H$ = 500 Oe, which is primarily the $ZFC$ curve. Then, to probe the memory effect we followed two different measurement paths. In path 1, the sample has been cooled to $T$ = 2 K and the measurement field of $H$ = 500 Oe is applied. We then record the $M-T$ response while warming. At some intermediate temperature which in this case is $T_m$ = 15 K, we reduced the field to $H$ = 300 Oe and wait for a time of $t_w$ = 1 hour. After 1 hour, we reapplied $H$ = 500 Oe and warm up the sample while recording the $M-T$ response. In path 2, we have performed a similar measurement procedure, but in this case instead of reducing the field to H = 300 Oe at $T_m$, we have rather increased the field by the same amount from H = 500 Oe to 700 Oe. It is interesting to note that in the first protocol, where the field has been reduced, the $M-T$ response (blue empty circles) overlaps with the reference curve and does not show any anomaly around $T_m$. This indicates that the system does not exhibit a memory effect while following the first measurement protocol. On the other hand, in the later protocol, when the field was increased, the $M-T$ response (red filled circle)  of the system shows a discontinuous jump in the magnetization at $T_m$ and on further warming, magnetization gradually merges with the reference magnetization curve at a slightly higher temperature than $T_m$. This clearly indicates an asymmetry on the effect of the magnetic field on the memory effect of the $ZFC$ state of the concentrated spin-glass systems. Earlier we have reported a similar asymmetrical memory effect in the spin-glass state {\color{blue}\cite{Sudip1}}, which underlines that this might be a generic feature of the metastable states of spin-glass materials.

%---------------------------------------------------------------------------------------------------------

\begin{figure}[t]
\centering
\includegraphics[scale=0.38]{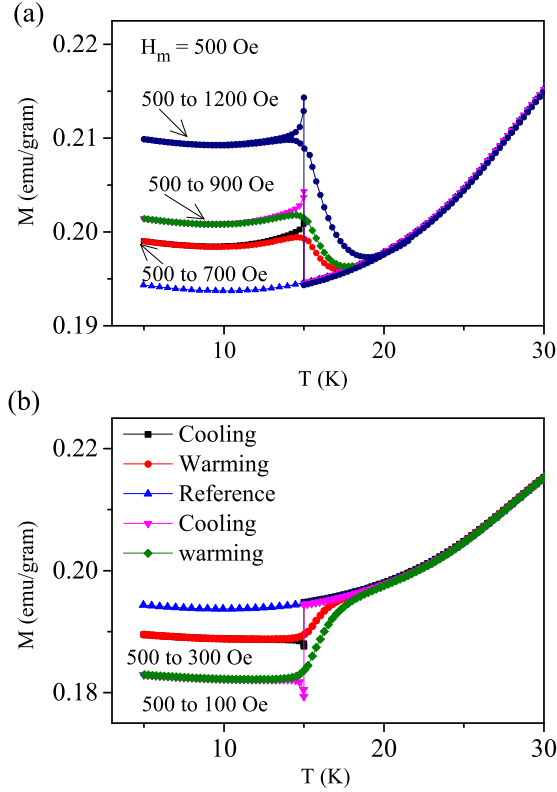}
\caption{Memory effect shown in the $FC$ state of AuFe(11\%) alloy when H is (a) increased and (b) decreased from the measuring field $H_m$ = 500 Oe.}
\end{figure}

%----------------------------------------------------------------------------------------------------------------------------------------------------------------------------------------------------------------------------
\subsection{Metastable field cooled state} 
In Fig. {\color{blue}4(a)}, {\color{blue}4(b)} and {\color{blue}4(c)}, we have shown the temperature variation of the ac susceptibility of AuFe(11\%) at an ac field of $H_{ac}$ = 1 Oe after cooling the sample across $T_f$ to 2 K in the presence of external dc magnetic field of $H_{dc}$ = 100, 500 and 1000 Oe, respectively.  At $H_{dc}$ = 100 Oe, it shows a cusp around $T_f$ = 33.5 K. As the measurement frequency increases, the cusp position does not change. All the curves are entirely merged above $T_f$, and has significant dispersion at lower temperatures. Let us define a characteristic temperature $T_d$, where the dispersion below $T_f$ starts. Note that, at $H_{dc}$ = 100 Oe, $T_f \sim T_d$.  At higher dc field $H_{dc}$ = 500 Oe, the cusp is rather diffused. It is interesting to note that the dispersion between ac susceptibility curves starts at lower temperature, i.e. $T_d < T_f$. As, we increase the dc bias field further, say to $H_{dc}$ = 1000 Oe, the cusp gets further broadened and the $T_d$ is shifted to further lower temperature. It indicates that in presence of dc field, $T_d$, where the dispersion between ac susceptibility at different frequencies begins is different from the $T_f$, and as the bias field $H_{dc}$ increases, the difference ($T_f - T_d$) progressively increases. Earlier we have shown the same frequency dpendence of ac susceptibility in the spin-glass state of dilute canonical spin-glasses $\cite{Sudip2}$ 

In Figs. {\color{blue}5(a)} and {\color{blue}5(b)}, we have presented the memory effect shown by the AuFe(11\%) in its FC state. In this case, initially the sample has been cooled from the paramagnetic state down to $T$ = 5 K in presence of the field $H_m$ = 500 Oe, and the temperature dependence of $M$ has been recorded at the same field in the warming cycle. Therefore, it is identical to the $FCW$ magnetization curve at H = 500 Oe, which we consider as the reference curve. In the next step, to check whether the sample shows memory effect in the field cooled state, we again measure the $M-T$ response of the sample following a different protocol. The sample is now cooled from the paramagnetic state down to temperature $T_m$ = 15 K at $H_m$ = 500 Oe. At, $T_m$, the cooling is stopped, and the field is raised from $H_m$ = 500 Oe to $H_i$ = 700 Oe for next $t_w$ = 1 hour. After the time period of $t_w$= 1 hour, $H_m$ = 500 Oe is reapplied and cooling is presumed down to the lowest temperature and subsequently warmed at the same field. In this case, the temperature variation of the $M$ is recorded in both cooling and heating cycles and it is compared with the reference curve to clearly reveal the memory effect. We have performed multiple measurements with the same measuring field of $H_m$ = 500 Oe and temperature $T_m$, but with different values of $H_i$ = 700, 900 and 1200 Oe.  In the Fig. {\color{blue}5(a)}, we have shown the temperature dependence of $M$ measured using this protocol and compared with the reference curve. If we carefully look at the $M-T$ curves at different values of $H_i$, a few distinct features can be observed:
\begin{enumerate}
\item In the cooling cycle, all magnetization curves are merged above $T_m$ and identical to the reference curve. When at $T_m$, field is cycled from $H_m \rightarrow H_i \rightarrow H_m$ with a waiting time of $t_w$ = 1 hour, $M$ does not return to the reference magnetization curve, and rather exhibits a step like jump at $T_m$. 
\item $M$ below $T_m$ in the cooling cycle depends on the value of $H_i$: larger the value of $H_i$, higher is the $M$.
\item Subsequently, when the sample is heated from the lowest temperature in the same magnetic field, $M-T$ curves merge with the reference curve at temperature say $T_{merge} > T_m$. Note that, $T_{merge}$ increases with the $H_i$.
\end{enumerate}   

The results presented above clearly indicate that the $FC$ state of the concentrated spin-glass system AuFe(11\%) is a non-equilibrium state and exhibits a distinct memory effect. The sample remembers its $FC$ state because even after the temporary rise or decrements of the magnetic field in the cooling cycle, the $M$ of the system returns to the previous value during warming. It underlines the complexity of the free energy landscape and non-equilibrium nature of the $FC$ state of the concentrated spin-glass systems. It may be noted here that apart from the mainstream mean-field theoretical models of spin-glass, a cluster model of a spin glass has also been considered sometimes for canonical spin-glass systems $\cite{klein,smith,levin}$. In this model, the long-range interactions between spins may lead to the formation of correlated spin clusters even above the spin freezing temperature. Such clusters form at a relatively higher temperature primarily due to the concentration fluctuation of impurity spins and they grow in size with decreasing temperature. Below the characteristic temperature $T_f$ there exist many finite clusters with varying size and shape. Within one cluster each spin is connected with at least one of the other spins in the same cluster. This kind of system has a broad distribution of the relaxation spectrum and therefore can be meaningful in explaining the observed non-equilibrium response of both the $FC$ and $ZFC$ states of AuFe(11\%) alloy.

%----------------------------------------------------------------------------------------------------------------------------------------------------------------------------------------------------------------------------
\section{Summary and conclusion} 
To summarize, we have performed magnetization and ac susceptibility measurements on the polycrystalline sample of AuFe(11\%) alloy. We probed both the $ZFC$ and $FC$ states in external magnetic fields up to $H$ = 70 kOe and temperatures down to $T$ = 2 K. In the $ZFC$ protocol at low fields, $M$ shows a distinct peak at $T_f$ followed by a characteristic temperature $T_{ir}$ slightly below $T_f$ where $ZFC$ and $FC$ magnetizations bifurcates giving rise to thermomagnetic irreversibility. A weak shoulder in $ZFC$ magnetization is then observed at a temperature $T_{Sh} < T_{ir}$, which becomes prominent with the increase in the dc magnetic field and marks the onset of a stronger thermomagnetic irreversibility. Thus, both the weak and strong irreversibility regimes as predicted by the mean-field theories of spin-glass and experimentally observed in other cluster spin-glasses, have been observed in the present AuFe(11\%) alloy. However, double logarithmic plots give an exponent $\alpha \approx$ 0.38 for both $T_{sh}-H$, and $T_{ir}-H$ lines, which is different from the expected mean-field values.  The thermomagnetic irreversibility ($\Delta M$) initially increases, shows a maximum, and then decreases with an increase in the external field. Interestingly, at higher fields $\Delta M$ reduces very slowly and even persists above H = 70 kOe. Both the $ZFC$ and $FC$ state of the cluster spin glass show memory effect in the dc magnetization measurements and frequency dependence in the ac-susceptibility measurements, thus highlighting the non-equilibrium nature of these states. This clearly suggests that the $FC$ state of a canonical cluster spin- AuFe(11\%) alloy also contains a rugged energy landscape, which is not consistent with the picture of a thermodynamic equilibrium state. This possibly indicates that  the cluster model $\cite{klein,smith,levin}$ for a spin glass may need some reconsideration to account for such experimental observations of metastable behavior.
%----------------------------------------------------------------------------------------------------------------------------------------------------------------------------------------------------------------------------
\section{Acknowledgment} 
S. B. Roy acknowledges financial support from Department of Atomic Energy, India in the form of Raja Ramanna Fellowship. Authors acknowledge Dr. Rajib Rawat at UGC DAE Consortium for scientific research, Indore, India for the SQUID-VSM facility to carry out magnetic measurements.
%----------------------------------------------------------------------------------------------------------------------------------------------------------------------------------------------------------------------------

\end{document}